\documentclass[12pt,a4paper,twoside,openany]{article}
\usepackage[dvips]{graphicx}
\begin{document}
\begin{titlepage}
\begin{center}
{\LARGE \bf PAIR FORMATION IN A t-J MODEL}
\end{center}
\vspace{1.0cm}
\begin{center}
{\large \bf  COURSE  PH-614}
\end{center}
\begin{center}
{\large \bf M.Sc. Project Report II}
\end{center}
\vspace{4.0cm}
\begin{center}
{\large \bf AYAN KHAN\footnote{akhan@iitg.ernet.in}\\Reg. No. 03212103\\M.Sc IVth. Semester}
\end{center}
\begin{center}                                                                  \hspace{8.0in}
\begin{figure}[!h]
\begin{center}
\includegraphics[width=10.0cm]{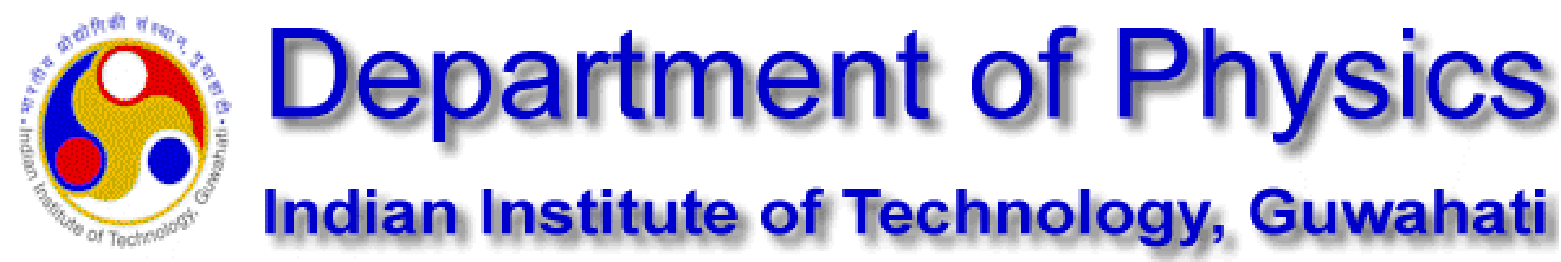}
\end{center}
\end{figure}
\end{center}
\vspace{3.0cm}
\begin{center}
{\large \bf Project Instructor : }
\end{center}
\begin{center}
{\large \bf Dr. SAURABH BASU\footnote{saurabh@iitg.ernet.in}\\Assistant Professor\\Department of Physics\\IIT Guwahati }
\end{center} 
\end{titlepage}
\newpage
\begin{center}
{\bf CERTIFICATE}
\end{center}
It is certified that the work contained in the project titled {\bf Pair Formation in a t-J Model} has been carried out by Ayan Khan, under my supervision.\\
\newline
\newline
\newline
\newline
{\bf Date:} 20. 04. 2005 \hspace{2.5 in}{\large \bf Dr. Saurabh Basu}\\\\
{\bf Place:} Guwahati    \hspace{2.60 in}{\bf Assistant Professor} \\

\hspace{3.68 in}{\bf Department of Physics}\\

\hspace{3.68 in}{\bf IIT Guwahati}
\newpage
\begin{center}
{\bf ACKNOWLEDGMENT}
\end{center}
I am deeply indebted to my project supervisor, Dr. Saurabh Basu, whose able 
guidence, thoughtful instruction invaluable criticism were instrumental in the 
progress of my work.\\\\ To my parents, I owe a special debt of gratitude for 
there blessings and support.\\\\ Above all I like to thank all my friends, my 
juniors and senior P.hD. scholars for their constant encouragement and invaluable suggetions.
\newline
\newline
\newline
\newline
{\bf Date:} 20/04/2005 \hspace{2.5 in}{\bf AYAN KHAN}
\newpage
\begin{titlepage}
\tableofcontents
\end{titlepage}
\newpage
\begin{center}
\setcounter{page}{1}\large \bf ABSTRACT
\end{center}
We have investigated the formation of bound state of two electrons in different
 kind of lattices using  a t-J-U model. In the model hopping parameter {\bf t} 
tries to delocalize the electrons where as pairing of electrons comes via 
Heisenberg exchange integral {\bf J} and hence it becomes necessary to 
calculate the threshold value of {\bf J}, viz. $J_{c}$, for which formation of 
bound states between two electron system is possible. The analysis is 
repeated for one dimensional chains, two dimensional square lattices, two leg 
ladders and three dimensional simple cubic lattices. Further we calculated the 
bound state energies for $J>J_{c}$. Also we have tried to shed some light 
to the symmetry operation of the lattices to understand the characteristic of 
two electron pairing. 
\newpage
\section{INTRODUCTION:}
The discovery of superconductivity in 1911 by Kammerlingh Onnes had given a new dimension in Condensed Matter Physics research. From the very beginning the scientific and commercial potential of superconductors had been well understood by the community. So as the days progressed different exciting features of superconductivity started to come to light and in 1987 with the discovery of high temperature superconductors the field of interest is further broadened.

Normal superconductors, which are usually good metals, are quite 
well understood by BCS theory where electron electron interaction is mainly 
controlled by phonons. But in high temperature superconductors the interaction picture among the electrons are still not clear.

Here our motivation is to study strongly correlated systems because 
it is well agreed that the origin of high temperature superconductivity is 
purely from electronic interaction, as for example here we are interested to 
study two dimensional square lattice which is analogous to the $CuO_{2}$ planes in a high temperature superconductor.

There are several 
models to study many particle systems and we are here using t-J model for studying electron pairing in metals with t denoting kinetic energy and J denoting Heisenberg exchange integral.We are starting with the 
assumption that the parent compounds are quite well represented by the 
antiferromagnetic Heisenberg model with localized electrons of spin $1/2$ 
occupying a lattice point and coupled by an exchange integral J. Doping is 
assumed to remove electrons thereby producing "holes" which are mobile 
because neighboring electrons can hop to the hole site with amplitude t. It 
has been shown earlier that \cite{Em} for the t-J model, dilute holes in a 
antiferromagnet are unstable against phase separation into a hole rich and a no
hole phase. It can argued that there exists a critical value of $J_{c}$,
such that when spin exchange integral J exceeds $J_{c}$, the hole rich phase
 has no electrons and for J slightly less than $J_{c}$ the hole rich phase is a 
low density superfluid of electron pairs.

Here we are investigating the critical
value of J for which pairing of electron is
possible in different kind of lattices. On the due course in the earlier
semester we had concentrated on one dimensional chains.
One dimensional analysis always carries a significant importance for its
relative simplicity, also if one dimensional lattice features are well
understood then it becomes relatively easy to understand the higher
dimensional lattices.
Then we had taken one step ahead to two dimensional square lattice. 
As we have stated the features of the two dimensional lattice has become very important after the discovery of high temperature superconductors in ceramic 
materials. The structural feature of ceramic materials is $CuO_{2}$ plane which are the main
source for high temperature superconductivity. So it is important to understand
 the electron correlation in two dimensional lattices.
Further we had extended our study on two leg ladder lattice. Since for the one-dimensional 
chain system quite a few things are known
exactly. One approach to tackle the superconducting cuprates is to investigate
the quasi one-dimensional lattices known as `spin ladder' structures,
which are strips of square lattice with a finite width and infinite length. An
example of spin ladder lattice system is $Sr_{n-1}Cu_{n+1}O_{2n}$
For these above mentioned lattices we had 
verified the critical value of J in square lattice which is {\bf 2t}\cite{Lin}.
Also we had investigated the critical value for one dimensional chain like 
lattice and two leg ladder where again we land up with the same result 
as $J=2t$. 

In this semester we were curious to look at the critical value of J 
in a simple cubic lattice. To understand the electronic interaction of the heavy-fermionic superconductors
such as $CeCu_{2}Si_{2}$, one needs to deal with different three dimensional
crystal structures. So it is a natural to look at a three dimensional
structure and try to find out the critical value of J for which pair formation
among two electrons is possible. So we calculated $J_{c}$ for a simple cubic lattice and this constitutes a new and central result in this project work. 
\section{MANY PARTICLE SYSTEM:}
\subsection{\bf Second Quantization:}
To understand the many body theory in condensed matter physics the essential 
technique is the method of second quantization. Soon after the foundation of 
quantum theory, the formalism of creation and annihilation operator (second 
quantization) was introduced. The physics of creation and annihilation 
operators can be explained in a better way from relativistic quantum field theory.
So second quantization is is nothing but a alternative formulation of quantum 
mechanics. the creation and annihilation operators are nothing but a tool that 
permits different process such as creation and annihilation of operators. Such 
process can not be discussed in the framework of Schr$\ddot{o}$dinger equation\cite{Second}.
\\We know that quantum mechanical wave function which represents a collection of
electrons is antisymmetric with respect to the operation which exchanges the 
space and spin coordinates of any two electrons. Thus, if $\psi(1, 2, 3,..., N)$
is an N electron wave function, and if $P_{ij}$ is the operator which exchanges
the coordinates if electron i and electron j, then $P_{ij}\Psi=-\Psi$\\J. C. 
Slater has introduced a method to represent such many electron wave functions. 
We need to begin with an orthonormal set of one-electron functions: $\phi_{1}, 
\phi_{2},....$, where $\int d\tau_{1}\phi_{i}^{\star}(1)\phi_{j}(1)=\delta_{ij}$\cite{Second}
, where 
\begin{displaymath}
\Delta=\frac{1}{\sqrt{N!}}\left |\begin{array}{cccc}
\phi_{1}(r_{1}) & ... & ... & \phi_{1}(r_{N})\\
... & ... & ... & ...\\
... & ... & ... & ...\\
\phi_{N}(r_{1}) & ... & ... & \phi_{N}(r_{N}) 
\end{array}\right|\equiv |\phi_{1}\phi_{2}\phi_{3}...|
\end{displaymath}
Such a function is called "Slater determinant." Since interchange of any two 
columns of a determinant changes its sign, $\Delta$ is antisymetric with respect
to the exchange operator $P_{ij}$.\\We can also represent antisymmetric many 
electron wave function in a different manner. Let us now define a set of 
electron creation operators, $b_{1}^{\dagger}, b_{2}^{\dagger},...,$, 
corresponding to the one electron spin orbitals, $\phi_{1}, \phi_{2}, ,...,$. 
When the creation operator $b_{j}^{\dagger}$ acts on an N-electron state, it produces an (N+1) electron state by creating an electron in the spin-orbital 
$\phi_{j}$. We used to denote no electron state or "vacuum state" as $|0>$. 
Similarly we define annihilation operator as $b_{j}$.\\ The commutation 
relations of the operators as follows:\\
\begin{center}
$b_{j}|0>=0$, $<0|b_{j}^{\dagger}=0$\\
$|b_{j}|N>=|N-1>$, $b_{j}^{\dagger}|N>=0$\\$b_{i}^{\dagger}b_{j}+b_{j}b_{i}^{\dagger}=\delta_{ij}$\\
$b_{i}^{\dagger}b_{j}^{\dagger}+b_{j}^{\dagger}b_{i}^{\dagger}=0$\\
$b_{i}b_{j}+b_{j}b_{i}=0$\\Now we can also represent Slater determinant with 
creation and annihilation operators, $\frac{1}{\sqrt{2}}\left |\begin{array}{cc}
\phi_{1}(r_{1}) & \phi_{1}(r_{2})\\
\phi_{2}(r_{1}) & \phi_{2}(r_{2})
\end{array}\right|\equiv|\phi_{1}\phi_{2}|=b_{i}^{\dagger}b_{j}^{\dagger}|0>$\\
Also $b_{i}^{\dagger}b_{j}^{\dagger}|0>=-b_{j}^{\dagger}b_{i}^{\dagger}|0>$\\
\end{center}
Now we can write kinetic energy operator as \\
\begin{displaymath}
\widehat{T}=\sum_{i,j}<i|T|j>b_{i}^{\dagger}b_{j}
\end{displaymath}
Similarly, the potential energy operator can be written as,
\begin{displaymath}
\widehat{V}=\frac{1}{2}\sum_{i,j,k,l}<ij|V({\mathbf r_{1},\mathbf r_{2}})|kl>b_{i}^{\dagger}b_{j}^{\dagger}b_{l}b_{k}
\end{displaymath}
Now if we define field operator as \\
\begin{displaymath}
\Psi({\mathbf r})=\sum_{i}b_{n_{i}}\phi_{n{i}}(\mathbf r)
\end{displaymath}
\begin{displaymath}
\Psi^{\dagger}({\mathbf r})=\sum_{i}b_{n_{i}}^{\star}\phi_{n{i}}^{\dagger}(\mathbf r)
\end{displaymath}
then 
\begin{displaymath}
<i|\widehat{T}|j>=\int{d^{3}r\phi_{n_{i}}}^{\dagger}(r)T(r)\phi_{n{i}}(r)
\end{displaymath}
\begin{displaymath}
<ij|\widehat{V}|kl>=\int{\int{d^{3}r_{1}d^{3}r_{2}\phi_{n_{i}}^{\dagger}(r_{1})\phi_{n_{j}}^{\dagger}(r_{2})}}V|r_{1}-r_{2}|\phi_{n_{i}}(r_{1})\phi_{n_{i}}(r_{2})
\end{displaymath}
If we consider Bloch wave function i.e $\phi({\mathbf r})=\frac{e^{i{\mathbf k}\cdot{\mathbf r}}}{\sqrt{V}}u_{kn}({\mathbf r})$, where $u_{kn}({\mathbf r})$ 
signifies particle is in periodic potential the Hamiltonian will be,
\begin{displaymath}
\widehat{H}=\sum_{k,\sigma}\epsilon_{k}b^{\dagger}_{k\sigma}b_{k\sigma}+\frac{1}{2}\sum_{q,\sigma,\sigma'}V_{q}b^{\dagger}_{k-q,\sigma}b_{k'+q,\sigma}b_{k'\sigma'}b_{k\sigma}
\end{displaymath}
this is second quantized Hamiltonian in many electron system.\\
As for example if we consider $\epsilon_{k}=-t\sum_{\sigma}e^{i{\mathbf k}\cdot{\mathbf{\delta}}}$, where t is hopping strength and $\delta$ is nearest neighbour where a electron
can hop. If our system is a two dimensional square lattice with lattice 
parameter {\bf a} then, $\delta=\pm{\widehat{x}a},\pm{\widehat{y}a}$, \\So our energy 
dispersion relation will be then \\
\begin{displaymath}
\epsilon_{k}=-t(e^{\pm ikx}+e^{\pm iky})
=-2t(\cos{k_{x}}+\cos{k_{y}})
\end{displaymath}
\subsection{\bf Theoretical Models in Many Particle Systems:}
\subsubsection{Tight Binding Model:}
In tight binding model the Hamiltonian describes the kinetic energy (hopping) 
of electrons for nearest neighbour pairs. 
\begin{displaymath}
\widehat{H}=-t\sum_{<i,j>,\sigma}(c_{i\sigma}^{\dagger}c_{j\sigma}+c_{j\sigma}^{\dagger}c_{i\sigma})
\end{displaymath}
In this approximation we consider the wave function of the electrons are sharply localized neglecting any overlap between them and they are confined in the 
lattice sites by an infinite potential barrier\cite{Mar}. From here on we like to fix the
 notation of creation and annihilation operator as $c_{i}$, $c_{j}^{\dagger}$ 
respectively.
\subsubsection{Heisenberg Model:}
The simplest model in quantum many body theory is isotropic spin half 
Heisenberg chain. The Heisenberg Hamiltonian is given by:
\begin{displaymath}
\widehat{H}=J{\sum_{<i,j>}({\mathbf{S}_{i}}{\mathbf{S}_{j}}-\frac{1}{4})}
\end{displaymath}
where $\mathbf{S}_{j}$ is a local spin variable at $j^{th.}$ state. For 
antiferromagnets $J>0$. The sum is over the distinct nearest neighbours. For 
spin 1/2 particle the spins are represented by Pauli's spin matrices.
\subsubsection{Hubbard Model:}
The Hubbard model describes the strongly correlated electron systems. The model
in more than one dimension has not been solved. In describing the $CuO_{2}$ 
planes in high temperature superconductivity Hubbard model is a good starting 
point\cite{Kam}.
The basic ingredients of Hubbard model are:\\
\begin{itemize}
\item{The kinetic energy (electron hopping) delocalizes the electron in Bloch 
state, leading to metallic behavior.}\cite{Fa}
\item{ The electron electron interaction (approximated by onsite Coulomb 
interaction) wants to localize the electron on to sites.}\cite{Fa}
\end{itemize}

The Hubbard model 
contains only one orbital per site and is defined as (considering nearest and 
next nearest neighbour interactions)\cite{Kam}:
\begin{displaymath}
\it \mathnormal{H} = -t\sum_{<i,j>,\sigma}(c_{i\sigma}^{\dagger}c_{j\sigma}+c_{j\sigma}^{\dagger}c_{i\sigma})-t'\sum_{<i,i'>,\sigma}(c_{i\sigma}^{\dagger}c_{i'\sigma}+c_{i'\sigma}^{\dagger}c_{i\sigma})+U \sum_{i} n_{i\uparrow}n_{i\downarrow}
\end{displaymath}
If we only take into account the nearest neighbour interaction it reduces to: 
\begin{displaymath}
\it \mathnormal{H} = -t\sum_{<i,j>,\sigma}(c_{i\sigma}^{\dagger}c_{j\sigma}+c_{j\sigma}^{\dagger}c_{i\sigma})+U \sum_{i} n_{i\uparrow}n_{i\downarrow}
\end{displaymath}
where $c_{i\sigma}^{\dagger}$ are creation operators and
$n_{i\sigma}=c_{i\sigma}^{\dagger}c_{i\sigma}$ are occupation number operator.
By means of on site Coulomb U the singlet band of the Hubbard model is split 
into a lower (LHB) and an upper Hubbard band (UHB). But 
the validity of the three band model to the single band model is still controversial \cite{Kam}. It has been questioned whether the strong coupling version of 
the Hamiltonian, i.e the t-J model, is appropriate to describe correctly the 
low energy physics of the original three band model. In the large U limit and 
at half filling (one electron per site) the Hubbard ladder is equivalent to the
spin 1/2 Heisenberg ladder \cite{Sup}.
\begin{displaymath}
\widehat{H}=J{\sum_{<i,j>}{\mathbf{S}_{i}}\cdot{\mathbf{S}_{j}}}
\end{displaymath}
When the Hubbard ladder is doped with holes away from half filling, its strong 
coupling description is modified from the Heisenberg model to the t-J model 
with the constrain of no doubly occupied sites\cite{Sup}.
\begin{displaymath}
\it \mathnormal{H} = -t\sum_{<i,j>,\sigma}(c_{i\sigma}^{\dagger}c_{j\sigma}+c_{j\sigma}^{\dagger}c_{i\sigma})+J{\sum_{<i,j>}({\mathbf{S}_{i}}\cdot{\mathbf{S}_{j}}-\frac{1}{4}n_{i}n_{j})}
\end{displaymath}
\subsubsection{t-J-U Model}
The most important local interactions in a doped antiferromagnet are well 
represented by the large U Hubbard model, the t-J model, and their various 
relatives. To be concrete we will focus on the t-J-U model. The t-J-U 
Hamiltonian is written as: 
\begin{displaymath}
\widehat{H} = -t \sum_{<i,j>}\sum_{\sigma} (c_{i\sigma}^{\dagger}c_{j\sigma}+c_{j\sigma}^{\dagger}c_{i\sigma})+J \sum_{<i,j>} ({\mathbf{S}_{i}}\cdot{\mathbf{S}_{j}} - \frac{n_{i}.n_{j}}{4})+U \sum_{i} n_{i\uparrow}n_{i\downarrow}
\end{displaymath}
It is a close variant of the familiar t-J model where the sites of the lattice 
is strictly prohibited against double occupancy, i.e the doubly occupied sites 
are projected out. The "no double occupancy" restriction is suitably achieved 
by using "constrained" fermionic operators. The same physics can be achieved by 
using a t-J-U model given by the limit $U\rightarrow\infty$ we go back to 
simple t-J model. The exchange integral J arises through virtual processes 
where in the intermediate state has a doubly occupied site, producing an 
antiferromagnetic coupling. Dopping is assumed to remove electrons thereby 
producing a "hole" or missing spin which is mobile because neighbouring 
electrons can hop into its place with amplitude t. So among the other models 
to study the correlation effects in the high temperature superconductors this 
model is simplest one and gives a exact critical value of $J(J_{c})$ such that $J \ge J_{c}$ two electrons can form a two particle bound state. 
\subsection{\bf Construction of Equation of Motion (EOM) for Two Particle System:}

The system consisting of two electrons the wave function can be written as
\begin{equation}
|\Psi\rangle=\sum_{i_{1},i_{2}}\Phi(i_{1},i_{2})c_{i_{1}\uparrow}^{\dagger}c_{i{2}\downarrow}^{\dagger}|0>
\end{equation}
and the model Hamiltonian is
\begin{equation}
\hat{H} = -t \sum_{<i,j>}\sum_{\sigma} (c_{i\sigma}^{\dagger}c_{j\sigma}+c_{j\sigma}^{\dagger}c_{i\sigma})+J \sum_{<i,j>} (S_{i}.S_{j} - \frac{n_{i}.n_{j}}{4})+U\sum_{i} n_{i\uparrow}n_{i\downarrow}
\end{equation}
where $|0\rangle$ denotes the vaccum state. For a two body problem the ground state isa singlet i.e $\Phi(i_{1},i_{2})=\Phi(i_{2}i_{1})$ and we know that \begin{math}H|\Psi\rangle=E|\Psi\rangle\end{math} so the equation of motion can be written as 
\begin{equation}
[-t\sum_{<i,j>}\sum_{\sigma} (c_{i\sigma}^{\dagger}c_{j\sigma}+c_{j\sigma}^{\dagger}c_{i\sigma})+J\sum_{<i,j>} (S_{i}.S_{j} - \frac{n_{i}.n_{j}}{4})+U\sum_{i} n_{i\uparrow}n_{i\downarrow}]\sum_{i_{1},i_{2}}\Phi(i_{1},i_{2})c_{i_{1}\uparrow}^{\dagger}c_{i{2}\downarrow}^{\dagger}|0>=E\sum_{i_{1},i_{2}}\Phi(i_{1},i_{2})
\end{equation}
\begin{displaymath}
\hat{t}|\Psi\rangle=-t\sum_{<i,j>}\sum_{\sigma} (c_{i\sigma}^{\dagger}c_{j\sigma}+c_{j\sigma}^{\dagger}c_{i\sigma})\sum_{i_{1},i_{2}}\Phi(i_{1},i_{2})c_{i_{1}\uparrow}^{\dagger}c_{i{2}\downarrow}^{\dagger}|0\rangle
\end{displaymath}
\begin{displaymath}
=-t\sum_{<i,j>}\sum_{i_{1},i_{2}}\Phi(i_{1},i_{2})(c_{i\uparrow}^{\dagger}c_{j\uparrow}c_{i_{1}\uparrow}^{\dagger}c_{i_{2}\downarrow}^{\dagger}|0\rangle+c_{j\uparrow}^{\dagger}c_{i\uparrow}c_{i_{1}\uparrow}^{\dagger}c_{i_{2}\downarrow}^{\dagger}|0\rangle+c_{i\downarrow}^{\dagger}c_{j\downarrow}c_{i_{1}\uparrow}^{\dagger}c_{i_{2}\downarrow}^{\dagger}|0\rangle+c_{j\downarrow}^{\dagger}c_{i\downarrow}c_{i_{1}\uparrow}^{\dagger}c_{i_{2}\downarrow}^{\dagger}|0\rangle)
\end{displaymath}
\begin{displaymath}
=\sum_{j}\Phi(j,i_{2})c_{i\uparrow}^{\dagger}c_{i_{2}\downarrow}^{\dagger}|0\rangle+0-\sum_{j}\Phi(i_{1},j)c_{i\downarrow}^{\dagger}c_{i_{1}\uparrow}^{\dagger}|0\rangle+0
\end{displaymath}
\begin{equation}
\hat{t}|\Psi\rangle=-t[\sum_{j}\Phi(j,i_{2})c_{i\uparrow}^{\dagger}c_{i_{2}\downarrow}|0\rangle-\Phi(i_{1},j)c_{i\downarrow}^{\dagger}c_{i_{1}\uparrow}^{\dagger}|0\rangle]
\end{equation}

\begin{displaymath}
\hat{U}|\Psi\rangle=U\sum_{<i,j>}\sum_{i_{1},i_{2}}\Phi(i_{1},i_{2})c_{i\uparrow}^{\dagger}c_{i\uparrow}c_{i\downarrow}^{\dagger}c_{i\downarrow}c_{i_{1}\uparrow}^{\dagger}c_{i_{2}\downarrow}^{\dagger}|0\rangle
\end{displaymath}
\begin{equation}
=U\sum_{i_{1},i_{2}}\Phi(i_{1},i_{2})\delta_{i_{1}i_{2}}c_{i_{1}\uparrow}^{\dagger}c_{i_{2}\downarrow}^{\dagger}|0\rangle
\end{equation}

\begin{displaymath}
\hat{J}|\Psi\rangle=J\sum_{<i,j>}\sum_{i_{1},i_{2}}\Phi(i_{1},i_{2})(\mathbf {S_{i}\cdot S_{j}}- \frac{n_{i}.n_{j}}{4})c_{i_{1}\uparrow}^{\dagger}c_{i_{2}\downarrow}^{\dagger}|0\rangle
\end{displaymath}
\begin{equation}
=J\sum_{i_{1},i_{2}}\Phi(i_{1},i_{2})c_{i_{1}\uparrow}^{\dagger}c_{i_{2}\downarrow}^{\dagger}|0\rangle
\end{equation}

\begin{equation}
E\Phi(i_{1},i_{2})=\sum_{j}[t_{i_{1}j}\Phi(j,i_{2})+t_{i_{2}j}\Phi(i_{1},j)]+[U\delta_{i_{1},i_{2}}-J_{i_{1},i_{2}}]\Phi(i_{1},i_{2})
\end{equation}
Fourier transform of the equation yields
\begin{equation}
E\Phi(\mathbf{k_{1},k_{2}})=[t(\mathbf{k_{1}})+t(\mathbf{k_{2}})]\Phi(\mathbf{k_{1},k_{2}})+\frac{U}{N}\sum_{k}\Phi(\mathbf{k_{1}+k,k_{2}-k})-\frac{1}{N}\sum_{k}J(\mathrm{k})\Phi(\mathbf{k_{1}-k,k_{2}+k})
\end{equation}
where,
\begin{displaymath}
\Phi(\mathbf{k_{1},k_{2}})=\frac{1}{N}\sum_{i_{1},i_{2}}\Phi(i_{1},i_{2})e^{-i\mathbf{k_{1}}\cdot\mathbf{r_{i_{1}}}-i\mathbf{k_{2}}\cdot\mathbf{r_{i_{2}}}}
\end{displaymath}
\begin{displaymath}
t(\mathbf{k})=\frac{1}{N}\sum_{i,j}t_{ij}e^{-\mathbf{k}\cdot(\mathbf{r_{i}}-\mathbf{r_{j}})}=-2t(\cos{k_{x}}+\cos{k_{y}})
\end{displaymath}
\begin{displaymath}
J(\mathbf{k})=2J(\cos{k_{x}}+\cos{k_{y}})
\end{displaymath}
taking the lattice constant 1.
Since the system is translationally invariant, the total momentum can be used to specify its eigenstates. let us define $\mathbf{Q}=\mathbf{k_{1}}+\mathbf{k_{2}}, \mathbf{q}=\frac{1}{2}(\mathbf{k_{1}}-\mathbf{k_{2}})$, and $\Phi(\mathbf{k_{1}},\mathbf{k_{2}})=\Phi_{Q}(\mathbf{q})$then we obtain
\begin{equation}
\Phi_{Q}(\mathbf{q})=\frac{\frac{U}{N}\sum_{k}\Phi_{Q}(\mathbf{k})-\frac{1}{N}\sum_{k}J(\mathbf{q}-\mathbf{k})\Phi_{Q}(\mathbf{k})}{E-t(\frac{\mathbf{Q}}{2}+\mathbf{q})-t(\frac{\mathbf{Q}}{2}-\mathbf{q})}
\end{equation}
This is the starting point of our analysis.
\section{CALCULATIONS:}
\subsection{\bf One Dimensional Chains:}
\begin{figure}[!h]
\begin{center}
\includegraphics[width=10.0cm]{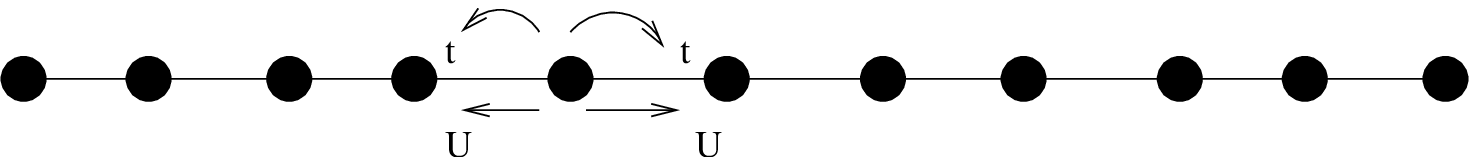}
\end{center}
\end{figure}
\begin{center}
Fig 1
\end{center}
For singlet pairing we can take Q=0 so from eq. (9) can be decoupled so that 
we can write,
\begin{equation}
C_{0}=UC_{0}I_{0}-2JI_{x}C_{x}
\end{equation}
\begin{equation}
C_{x}=UC_{0}I_{x}-2JI_{xx}C_{x}
\end{equation}
where 
\begin{displaymath}
C_{0}=\frac{1}{N}\sum_{\mathbf{q}}\Phi_{0}(\mathbf{q})
\end{displaymath}
\begin{displaymath}
C_{x}=\frac{1}{N}\sum_{k}\cos{k_{x}}\Phi_{0}(\mathbf{k})
\end{displaymath}
\begin{displaymath}
I_{0}=\frac{1}{N}\sum_{q}{\frac{1}{E+4t\cos{q_{x}}}}
\end{displaymath}
\begin{displaymath}
I_{x}=\frac{1}{N}\sum_{q}{\frac{\cos{q_{x}}}{E+4t\cos{q_{x}}}}
\end{displaymath}
\begin{displaymath}
I_{xx}=\frac{1}{N}\sum_{q}{\frac{\cos^{2}{q_{x}}}{E+4t\cos{q_{x}}}}
\end{displaymath}
eq. (11) and (12) can be written in matrix form as follows:
\begin{equation}
{\left (\begin{array}{cc}
UI_{0}-1 & -2JI_{x}\\
UI_{x} & -2JI_{xx}-1
\end{array}\right)}
{\left(\begin{array}{c}
C_{0}\\C_{x}
\end{array}\right)}=0
\end{equation}
In eq. (12) unique solution of $C_{0}$ and $C_{x}$ will exit if and only if the determinant of the coefficient is zero.
\begin{equation}
\left |\begin{array}{cc}
UI_{0}-1 & -2JI_{x}\\
UI_{x} & -2JI_{xx}-1
\end{array}\right|=0
\end{equation}
Now solving eq. (13) for J 
\begin{equation}
-2J=\frac{1-UI_{0}}{I^2_{x}U+I_{xx}(1-UI_{0})}
\end{equation}
Using the lattice symmetry we can write \\
\begin{displaymath}
I_{0}=-\frac{1}{4}\frac{1}{\sqrt{\alpha^2-1}}
\end{displaymath}
\begin{displaymath}
I_{x}=\frac{1}{4}\frac{\alpha+\sqrt{\alpha^2-1}}{\sqrt{\alpha^2-1}}
\end{displaymath}
\begin{displaymath}
I_{xx}=-\frac{1}{4}\frac{\alpha(\alpha+\sqrt{\alpha^2-1})}{\sqrt{\alpha^2-1}}
\end{displaymath}
eq. (15), (16), (17) can be written is a more simple form as
\begin{equation}
I_{x}=\frac{1}{4t}-\frac{E}{4t}I_{0} 
\end{equation}
\begin{equation}
I_{xx}=-\frac{E}{4t}I_{x}
\end{equation}
substituting these values with the limit as $I_{0}\rightarrow\infty$ and $U\rightarrow\infty$ in eq. (14) we get $\bf J_{c}=2t$\\

{\bf Bound state energy of electron in one dimension}\\
If we think in terms of the energy band in one dimension it is 8t for our system, so the bound state energy E of the two particle system can be obtained via equation (14), (15), (16) is we can write
\begin{equation}
\frac{8t}{J}=\frac{1}{tI_{0}}-\frac{E}{t}
\end{equation}
Taking $t=1$ we can write the energy equation as 
\begin{equation}
E=\frac{1}{I_{0}}-\frac{8t^{2}}{J}
\end{equation}
\newpage
\begin{figure}[!h]
\begin{center}
\includegraphics[width=12.0cm,angle=270]{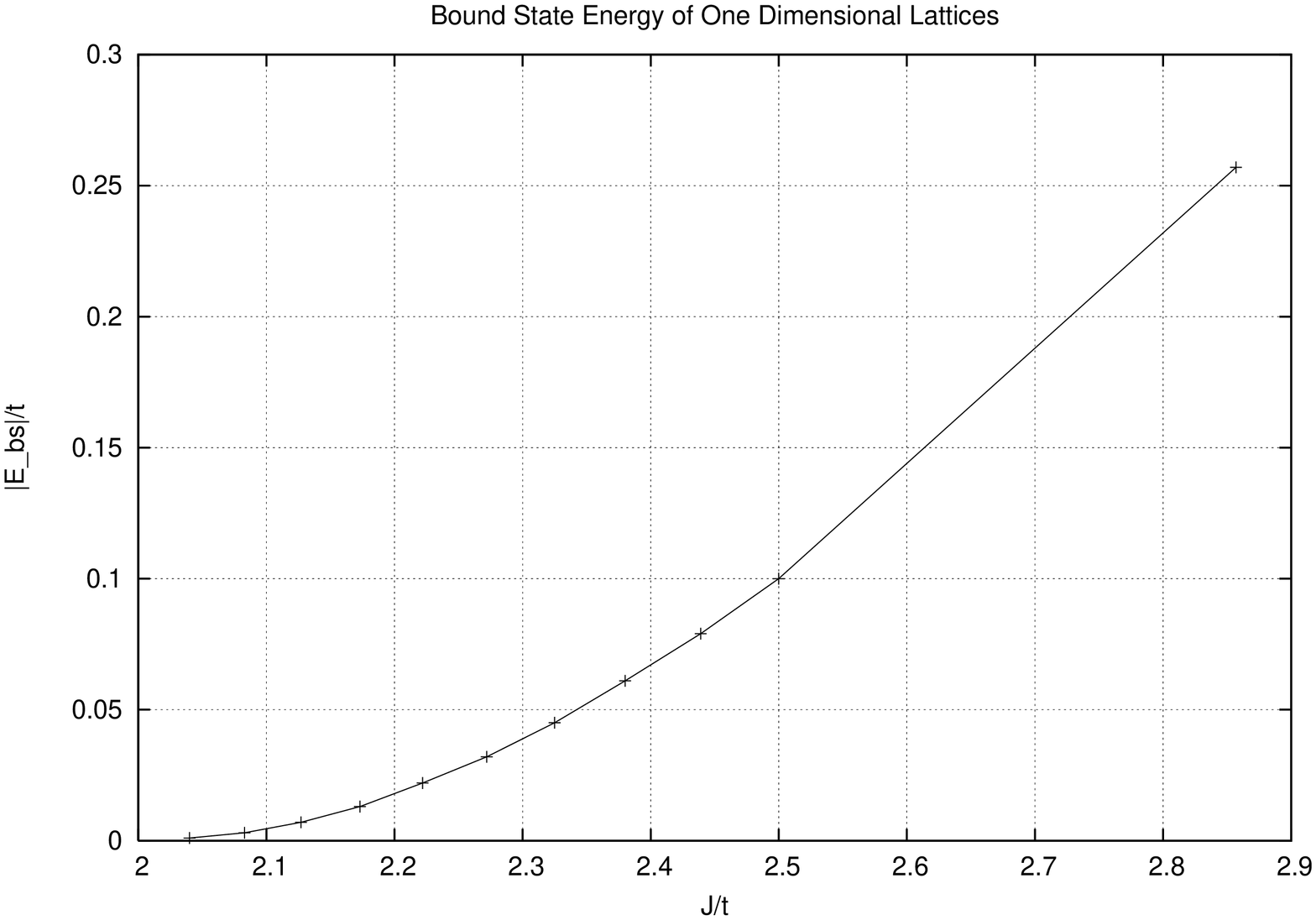}
\end{center}
\end{figure}
\begin{center}
Fig 2
\end{center}
\newpage
\subsection{\bf Two Dimensional Square Lattice:}
\begin{figure}[!h]
\begin{center}
\includegraphics[width=10.0cm]{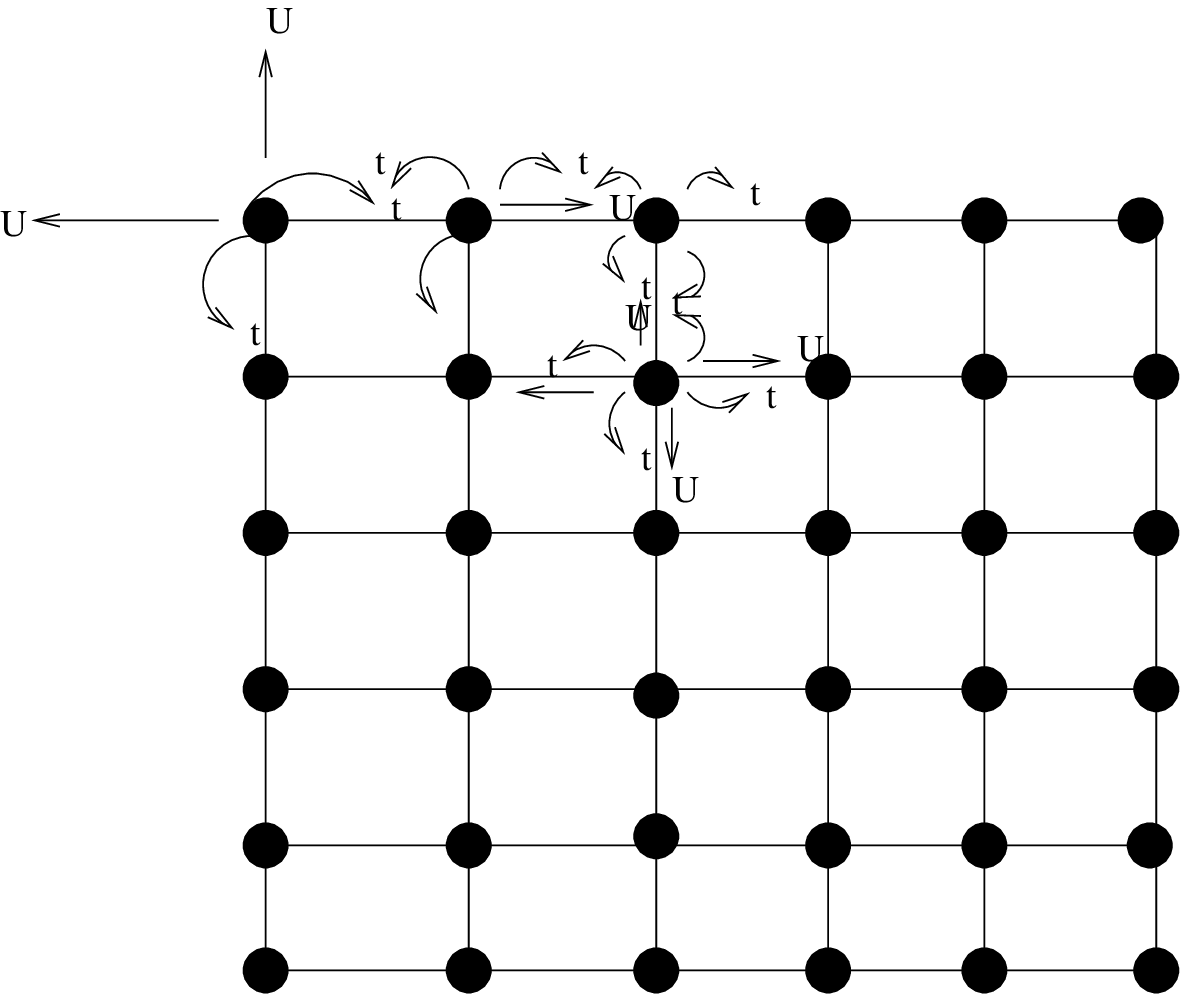}
\end{center}
\end{figure}
\begin{center}
Fig 3
\end{center}
In a similar fashion as we have done in the previous section we can write the following equations:
\begin{equation}
C_{0}=UI_{0}C_{0}-2JI_{x}C_{x}-2JI_{y}C_{y}
\end{equation}
\begin{equation}
C_{x}=UC_{0}I_{x}-2JC_{x}I_{xx}-2JC_{y}I_{xy}
\end{equation}
\begin{equation}
C_{y}=UC_{0}I_{y}-2JC_{x}I_{xy}-2JC_{y}I_{yy}
\end{equation}
where
\begin{displaymath}
C_{0}=\frac{1}{N}\sum_{\mathbf{q}}\Phi_{0}(\mathbf{q})
\end{displaymath}
\begin{displaymath}
C_{x}=\frac{1}{N}\sum_{k}\cos{k_{x}}\Phi_{0}(\mathbf{k})
\end{displaymath}
\begin{displaymath}
C_{y}=\frac{1}{N}\sum_{k}\cos{k_{y}}\Phi_{0}(\mathbf{k})
\end{displaymath}
\begin{displaymath}
I_{0}=\frac{1}{N}\sum_{q}{\frac{1}{E+4t(\cos{q_{x}+\cos{q_{y}})}}}
\end{displaymath}
\begin{displaymath}
I_{x}=\frac{1}{N}\sum_{q}{\frac{\cos{q_{x}}}{E+4t(\cos{q_{x}+\cos{q_{y}})}}}
\end{displaymath}
\begin{displaymath}
I_{xx}=\frac{1}{N}\sum_{q}{\frac{\cos^{2}{q_{x}}}{E+4t\cos{q_{x}}}}
\end{displaymath}
\begin{displaymath}
I_{y}=\frac{1}{N}\sum_{q}{\frac{\cos{q_{y}}}{E+4t(\cos{q_{x}+\cos{q_{y}})}}}
\end{displaymath}
\begin{displaymath}
I_{xy}=\frac{1}{N}\sum_{q}{\frac{\cos{q_{x}}\cos{q_{y}}}{E+4t(\cos{q_{x}}+\cos{q_{y}})}}
\end{displaymath}
\begin{displaymath}
I_{yy}=\frac{1}{N}\sum_{q}{\frac{\cos^{2}{q_{y}}}{E+4t(\cos{q_{x}}+\cos{q_{y}})}}
\end{displaymath}
eq. (20),(21),(22) can be written in a matrix form as follows:
\begin{equation}
{\left (\begin{array}{ccc}
UI_{0}-1 & -2JI_{x} & -2JI_{y}\\
UI_{x} & -2JI_{xx}-1 & -2JI_{xy}\\
UI_{y} & -2JI_{xy} & -2JI_{yy}-1
\end{array}\right)}
{\left(\begin{array}{c}
C_{0}\\C_{x}\\C_{y}
\end{array}\right)}=0
\end{equation}
In eq. (23) unique solution of $C_{0}$, $C_{x}$ and $C_{y}$ will exit if and only if the determinant of the coefficient matrix is zero.
\begin{equation}
\left |\begin{array}{ccc}
UI_{0}-1 & -2JI_{x} & -2JI_{y}\\
UI_{x} & -2JI_{xx}-1 & -2JI_{xy}\\
UI_{y} & -2JI_{xy} & -2JI_{yy}-1
\end{array}\right|=0
\end{equation}
For an isotropic square lattice symmetry permits us to write $I_{y}=I_{x}$, $I_{yy}=I_{xx}$ and $C_{y}=C_{x}$
hence eq. (23) becomes
\begin{equation}
{\left (\begin{array}{ccc}
UI_{0}-1 & -2JI_{x} & -2JI_{x}\\
UI_{x} & -2JI_{xx}-1 & -2JI_{xy}\\
UI_{y} & -2JI_{xy} & -2JI_{xx}-1
\end{array}\right)}
{\left(\begin{array}{c}
C_{0}\\C_{x}\\C_{x}
\end{array}\right)}=0
\end{equation}
and our modified determinant will be,
\begin{equation}
\left |\begin{array}{ccc}
UI_{0}-1 & -2JI_{x} & -2JI_{x}\\
UI_{x} & -2JI_{xx}-1 & -2JI_{xy}\\
UI_{x} & -2JI_{xy} & -2JI_{xx}-1
\end{array}\right|=0
\end{equation}
Now our motivation is to take $U\rightarrow\infty$ limit to project out the possibility of double occupancy. The value of $I_{0}, I_{x}, I_{xx}, I_{xy}$ after integration are as follows:
\begin{equation}
I_{0}=\frac{1}{2}\frac{K(\frac{-2}{\alpha})}{\pi\alpha}
\end{equation}
\begin{equation}
I_{x}=\frac{1}{2}\frac{(1-\alpha)K(\frac{-2}{\alpha})}{\pi\alpha}-\frac{1}{2}\frac{(\alpha-2)\Pi(\frac{2}{\alpha},\frac{-2}{\alpha})}{\pi}
\end{equation}
\begin{equation}
I_{xx}=\frac{1}{4}\frac{({\alpha}^2-2\alpha+2)K(\frac{-2}{\alpha})}{\pi\alpha}-\frac{1}{2}\frac{(\alpha-2)\Pi(\frac{2}{\alpha},\frac{-2}{\alpha})}{\pi}+\frac{1}{4}\frac{{\alpha}E(-\frac{2}{\alpha})}{\pi}
\end{equation}
\begin{equation}
I_{xy}=-\frac{1}{2}\frac{(1-\alpha)K(\frac{-2}{\alpha})}{\pi}-\frac{1}{2}\frac{(\frac{1}{2}\alpha-\alpha+1)K(-\frac{2}{\alpha})}{\pi\alpha}-\frac{1}{4}\frac{{\alpha}E(-\frac{2}{\alpha})}{\pi}
\end{equation}
These expression can be written in a simple form as:\\
\begin{equation}
I_{xx}+I_{xy}=-\frac{E}{4t}I_{x}
\end{equation} 
\begin{equation}
I_{x}=\frac{1}{8t}-\frac{E}{8t}I_{0}
\end{equation}
\begin{equation}
I_{0}=\frac{1}{E}\frac{2}{\pi}K(\frac{8t}{E})
\end{equation}
So the lattice integrals are turned out in terms of complete elliptic integral 
of first kind ($K(k)$), second kind ($E(k)$), third kind ($\Pi(\nu,k)$) 
respectively. $\Pi(\nu,k)$ can be evaluated through $K(k)$, $E(k)$, $F(q,\phi)$,
 $E(q,\phi)$, where $F(q,\phi)$ and  $E(q,\phi)$ are the incomplete Elliptic 
integrals of first and second kind respectively\cite{Jer}. The nature of $K(k)$ and  $E(k)$
is shown as follows:\\
\begin{figure}[!h]
\begin{center}
\includegraphics[width=18.0cm, height=7.0cm]{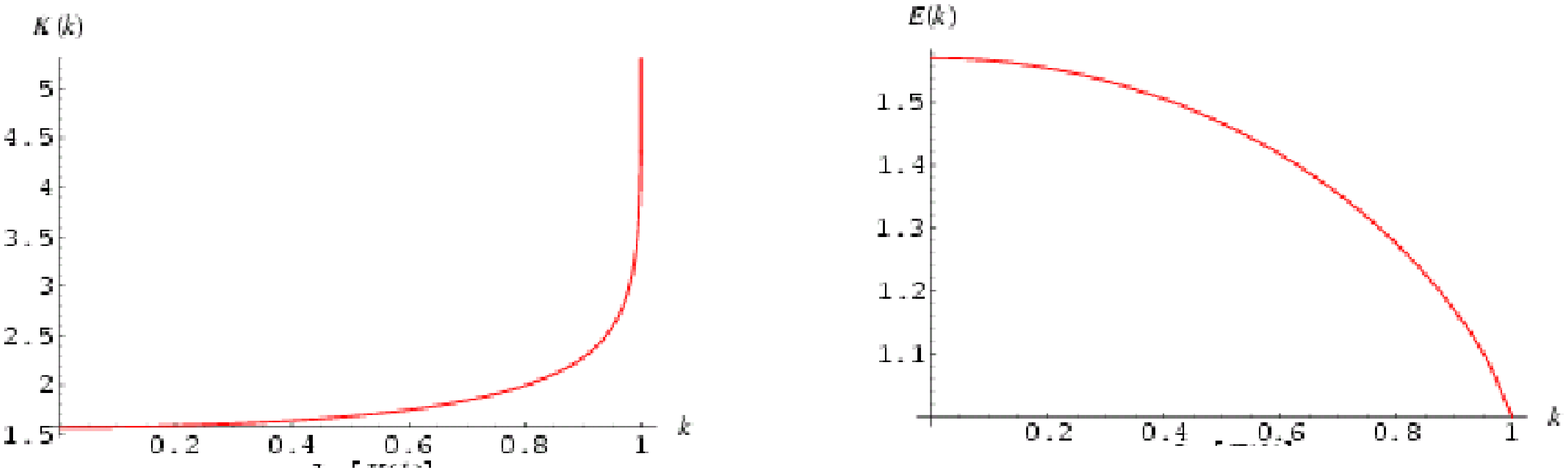}
\end{center}
\end{figure}
\begin{center}
Fig 4
\end{center}
So from the physical nature of the elliptic integral of the first kind we can 
conclude that it is of logarithmic diverging nature so to 
tackle this problem let us consider $\alpha=-2-\delta$ and expand the elliptic 
integrals w.r.t $\delta$ then take $\delta\rightarrow0$ limit. Since the 
determinant is zero thus the coefficient of the diverging $ln\delta$ term 
should be equal to zero.\\ 
So $-\frac{1}{32}\frac{J^2}{\pi}+{1}{8}\frac{1}{\pi}-\frac{1}{4}\frac{J}{\pi^2}+\frac{1}{8}\frac{J^2}{\pi^2}=0$\\The solution of the quadratic equation for 
{\bf J= 2t} and {\bf 7.32t}.\\   

{\bf Bound state energy of electron in two dimension}\\
From eq. (24) we can write 
\begin{equation}
-2J=\frac{1-UI_{0}}{2UI_{x}^2+(I_{xx}+I_{xy})(1-UI_{0})}
\end{equation}
From eq. (31), (32), (33), and (34) the bound state energy E of the two electron system can be obtained via 
\begin{equation}
\frac{16t}{J}=\frac{\pi}{2}\frac{\frac{E}{t}}{K(\frac{8t}{E})}-\frac{E}{t}
\end{equation}
Taking $t=1$ we rewrite the equation as:
\begin{equation}
E=\frac{16}{J(\frac{2}{\pi}\frac{1}{K(\alpha)}-1)}
\end{equation}
\begin{figure}[!h]
\begin{center}
\includegraphics[width=12.0cm,angle=270]{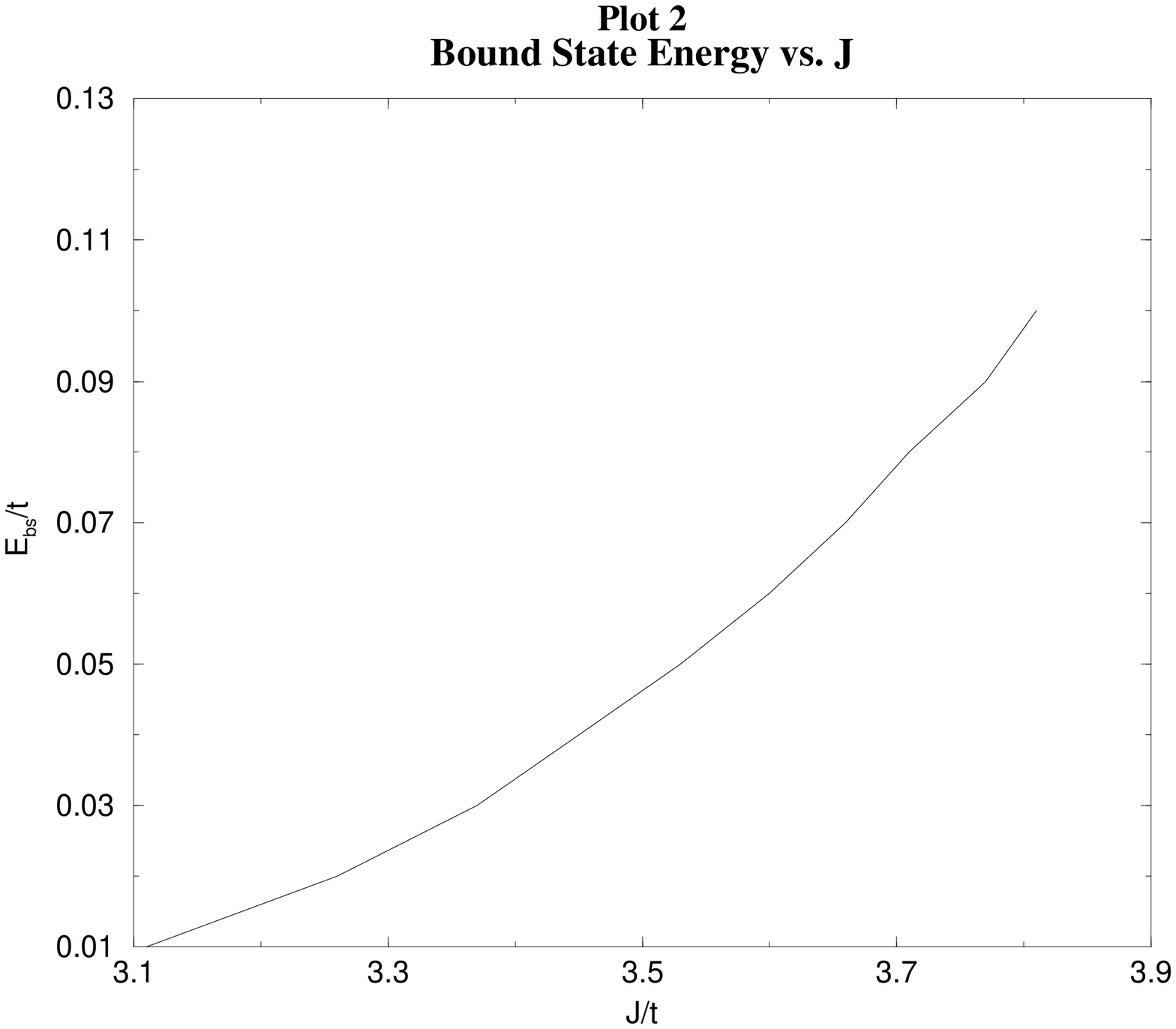}
\end{center}
\end{figure}
\begin{center}
Fig 5
\end{center}
\newpage
\subsection{\bf Two Leg Ladder:}
\begin{figure}[!h]
\begin{center}
\includegraphics[width=10.0cm]{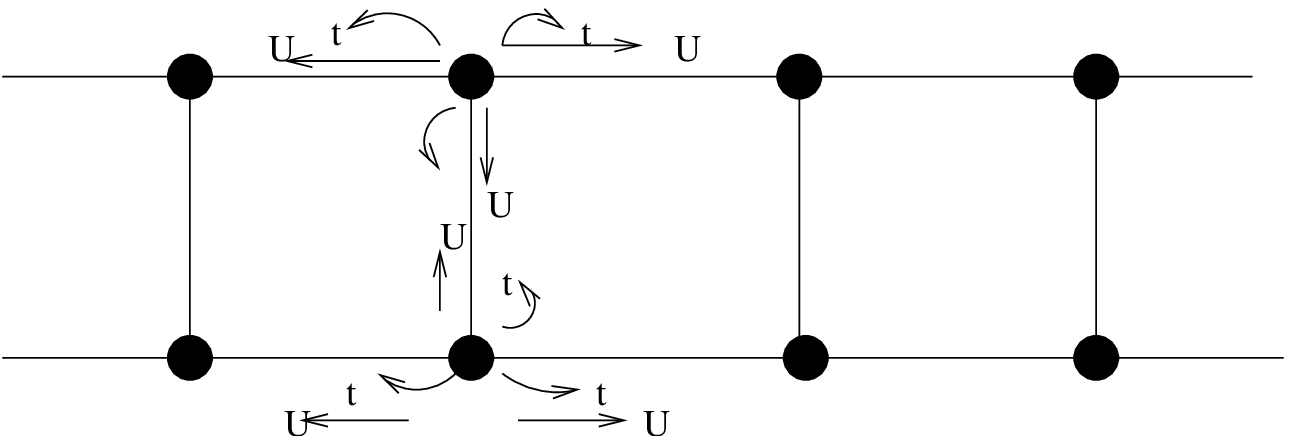}
\end{center}
\end{figure}
\begin{center}
Fig 6
\end{center}
So a ladder like lattice involves a one dimensional wave vector integral 
({\it viz} over $q_{x}$) rather that two dimensional integral (over $q_{x}$ and
$ q_{y}$). More precisely the lattice integrals appearing in the calculation of bound states are expressed as,
\begin{displaymath}
\sum_{q}=\frac{1}{2}\sum_{q_{y}=0,2\pi}\frac{1}{2\pi}\int_{-\pi}^{\pi}dq_{x}
\end{displaymath} 
The various lattice integrals are as follows:
\begin{equation}
I_{0}=\frac{1}{4}\int_{-\pi}^{\pi}\frac{dx}{\alpha+1+\cos{q_{x}}}+\frac{1}{4}\int_{-\pi}^{\pi}\frac{dx}{\alpha-1+\cos{q_{x}}}
\end{equation}
\begin{equation}
I_{x}=\frac{1}{4}\int_{-\pi}^{\pi}\frac{\cos{q_{x}}dx}{\alpha+1+\cos{q_{x}}}+\frac{1}{4}\int_{-\pi}^{\pi}\frac{\cos{q_{x}}dx}{\alpha-1+\cos{q_{x}}}
\end{equation}
\begin{equation}
I_{xx}=\frac{1}{4}\int_{-\pi}^{\pi}\frac{\cos^{2}{q_{x}}dx}{\alpha+1+\cos{q_{x}}}+\frac{1}{4}\int_{-\pi}^{\pi}\frac{\cos^{2}{q_{x}}dx}{\alpha-1+\cos{q_{x}}}
\end{equation}
where $\alpha=\frac{E}{4t}$\\from eq. (13) with $t=1$ and $U\rightarrow\infty$
we can write, 
\begin{equation}
\left |\begin{array}{cc}
I_{0} & -2JI_{x}\\
I_{x} & -2JI_{xx}-1
\end{array}\right|=0
\end{equation}
expanding the determinant we obtain
\begin{equation}
2JI_{x}^2-2JI_{0}I_{xx}-I_{0}=0
\end{equation}
We have substituted the integral values in eq. (42) and also $\alpha$ is replaced by $=-\delta-2$. Then expanding the entire equation  w.r.t $\delta$ we got a term of $\frac{1}{\sqrt{\delta}}$. This term will diverge as soon as we will take the limit as $\delta\rightarrow0$. So collecting the coefficients of this term and equating them to zero the critical value of J is obtained which is 
{\bf 2t}.\\

{\bf Bound state energy of Two leg ladder}\\
Directly from the integrals one can find out the bound state energy for 
pairing of electrons with proper substitution. From the plot also we can verify
 that the minimum energy required for formation of bound state among two 
electrons in a ladder like lattice is {\bf 2}, considering t=1.
\newpage
\begin{figure}[!h]
\begin{center}
\includegraphics[width=12.0cm,angle=270]{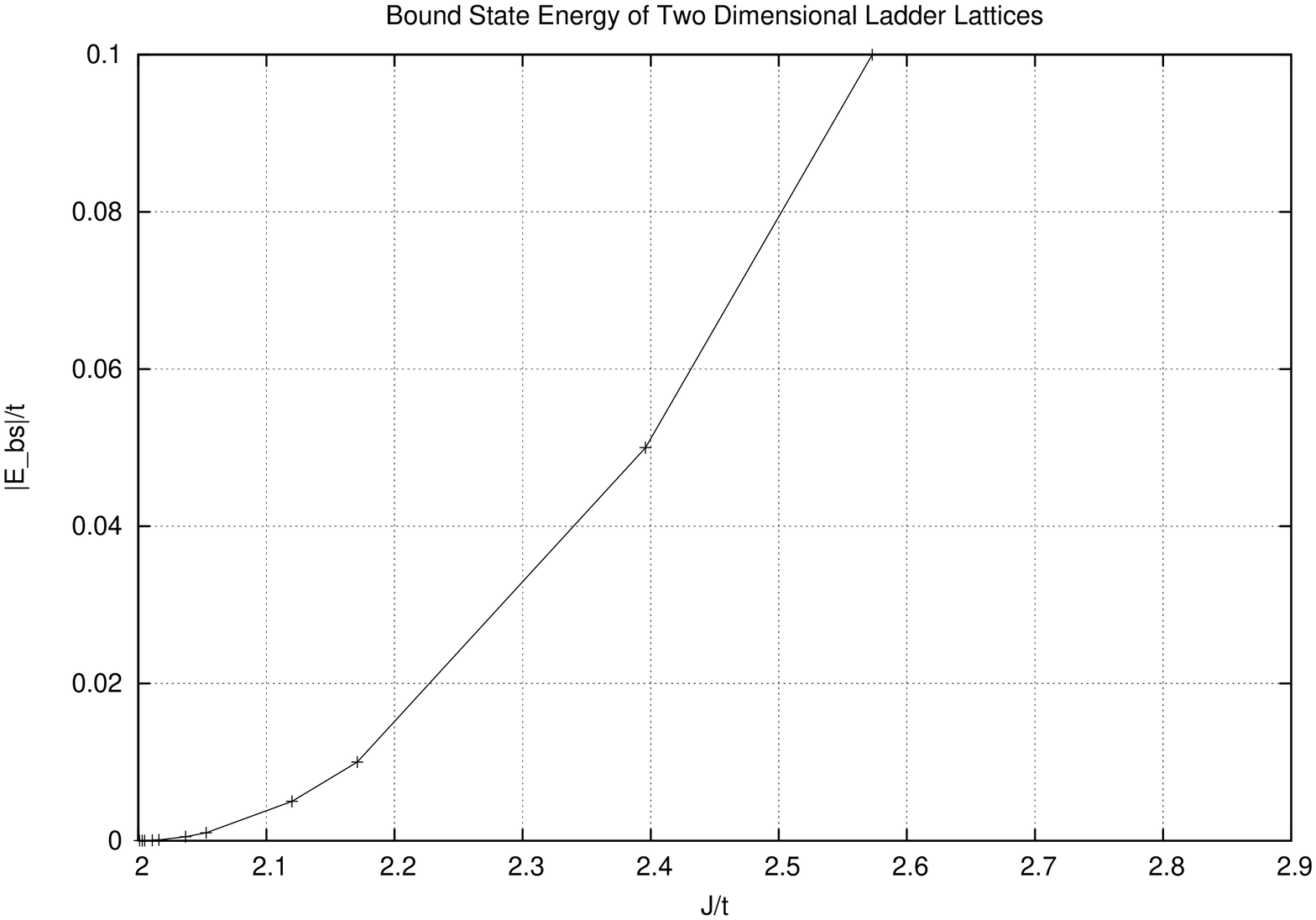}
\end{center}
\end{figure}
\begin{center}
Fig 7
\end{center}
\newpage
\subsection{\bf Three Dimensional Cubic Lattice:}
\begin{figure}[!h]
\begin{center}
\includegraphics[width=10.0cm]{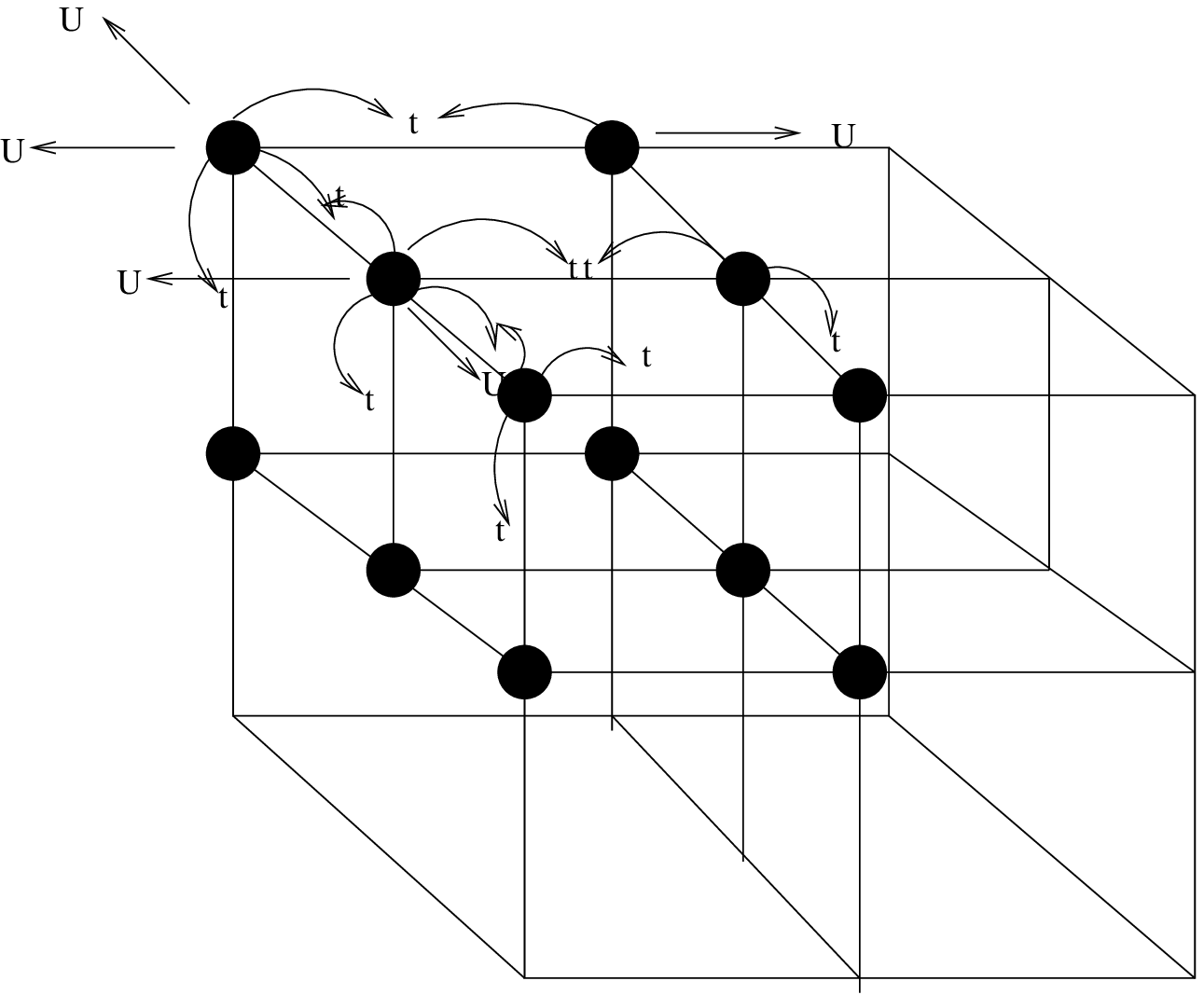}
\end{center}
\end{figure}
\begin{center}
Fig 8
\end{center}
From eq.(9) considering that the lattice is isotropic (Q=0), we can write
\begin{equation}
\Phi_{0}(\mathbf{q})=\frac{\frac{U}{N}\sum_{k}\Phi_{0}(\mathbf{k})-\frac{1}{N}\sum_{k}J(\mathbf{q}-\mathbf{k})\Phi_{0}(\mathbf{k})}{E-2t(\mathbf{q})}
\end{equation} 
\begin{equation}
\Phi_{0}(\mathbf{q})=\frac{\frac{U}{N}\sum_{k}\Phi_{0}(\mathbf{k})-\frac{1}{N}\sum_{k}J(\mathbf{q}-\mathbf{k})\Phi_{0}(\mathbf{k})}{E+4t(\cos{q_{x}}+\cos{q_{y}}+\cos{q_{z}})}
\end{equation}
This equation can be written as follows:
\begin{equation}
C_{0}=UC_{0}I_{0}-2JC_{x}I_{x}-2JC_{y}I_{y}-2JC_{z}I_{z}
\end{equation}
\begin{equation}
C_{x}=UC_{0}I_{x}-2JC_{x}I_{xx}-2JC_{y}I_{xy}-2JC_{z}I_{xz}
\end{equation}
\begin{equation}
C_{y}=UC_{0}I_{y}-2JC_{x}I_{yx}-2JC_{y}I_{yy}-2JC_{z}I_{yz}
\end{equation}
\begin{equation}
C_{z}=UC_{0}I_{z}-2JC_{x}I_{zx}-2JC_{y}I_{zy}-2JC_{z}I_{zz}
\end{equation}
where
\begin{displaymath}
C_{0}=\frac{1}{N}\sum_{\mathbf{q}}\Phi_{0}(\mathbf{q})
\end{displaymath}
\begin{displaymath}
C_{x}=\frac{1}{N}\sum_{k}\cos{k_{x}}\Phi_{0}(\mathbf{k})
\end{displaymath}
\begin{displaymath}
C_{y}=\frac{1}{N}\sum_{k}\cos{k_{y}}\Phi_{0}(\mathbf{k})
\end{displaymath}
\begin{displaymath}
C_{z}=\frac{1}{N}\sum_{k}\cos{k_{z}}\Phi_{0}(\mathbf{k})
\end{displaymath}
\begin{displaymath}
I_{0}=\frac{1}{N}\sum_{q}{\frac{1}{E+4t(\cos{q_{x}+\cos{q_{y}}+\cos{q_{z}})}}}
\end{displaymath}
\begin{displaymath}
I_{x}=\frac{1}{N}\sum_{q}{\frac{\cos{q_{x}}}{E+4t(\cos{q_{x}+\cos{q_{y}}+\cos{q_{z}})}}}
\end{displaymath}
\begin{displaymath}
I_{xx}=\frac{1}{N}\sum_{q}{\frac{\cos^{2}{q_{x}}}{E+4t(\cos{q_{x}}+\cos{q_{y}}+\cos{q_{z}})}}
\end{displaymath}
\begin{displaymath}
I_{y}=\frac{1}{N}\sum{q}{\frac{\cos{q_{y}}}{E+4t(\cos{q_{x}+\cos{q_{y}}+\cos{q_{z}})}}}
\end{displaymath}
\begin{displaymath}
I_{xy}=\frac{1}{N}\sum_{q}{\frac{\cos{q_{x}}\cos{q_{y}}}{E+4t(\cos{q_{x}}+\cos{q_{y}}+\cos{q_{z}})}}
\end{displaymath}
\begin{displaymath}
I_{yy}=\frac{1}{N}\sum_{q}{\frac{\cos^{2}{q_{x}}}{E+4t(\cos{q_{x}}+\cos{q_{y}}+\cos{q_{z}})}}
\end{displaymath}
\begin{displaymath}
I_{z}=\frac{1}{N}\sum_{q}{\frac{\cos{q_{z}}}{E+4t(\cos{q_{x}+\cos{q_{y}}+\cos{q_{z}})}}}
\end{displaymath}
\begin{displaymath}
I_{zz}=\frac{1}{N}\sum_{q}{\frac{\cos^{2}{q_{z}}}{E+4t(\cos{q_{x}}+\cos{q_{y}}+\cos{q_{z}})}}
\end{displaymath}
\begin{displaymath}
I_{zx}=\frac{1}{N}\sum_{q}{\frac{\cos{q_{z}}\cos{q_{x}}}{E+4t(\cos{q_{x}}+\cos{q_{y}}+\cos{q_{z}})}}
\end{displaymath}
\begin{displaymath}
I_{yz}=\frac{1}{N}\sum_{q}{\frac{\cos{q_{y}}\cos{q_{z}}}{E+4t(\cos{q_{x}}+\cos{q_{y}}+\cos{q_{z}})}}
\end{displaymath}
eq. (44), (45), (46), (47) can be written in a matrix form as follows:
\begin{equation}
{\left (\begin{array}{cccc}
UI_{0}-1 & -2JI_{x} & -2JI_{y} & -2JI_{z}\\
UI_{x} & -2JI_{xx}-1 & -2JI_{xy} & -2JI_{zx}\\
UI_{y} & -2JI_{xy} & -2JI_{yy}-1 & -2JI_{yz}\\
UI_{z} & -2JI_{zx} & -2JI_{yz} & -2JI_{zz}-1
\end{array}\right)}
{\left(\begin{array}{c}
C_{0}\\C_{x}\\C_{y}\\C_{z}
\end{array}\right)}=0
\end{equation}
$J_{c}/t$ can be found out by vanishing of the determinant by choosing an energy E to be slightly between noninteracting 2 electron band.

\begin{displaymath}
{\left |\begin{array}{cccc}
UI_{0}-1 & -2JI_{x} & -2JI_{y} & -2JI_{z}\\
UI_{x} & -2JI_{xx}-1 & -2JI_{xy} & -2JI_{zx}\\
UI_{y} & -2JI_{xy} & -2JI_{yy}-1 & -2JI_{yz}\\
UI_{z} & -2JI_{zx} & -2JI_{yz} & -2JI_{zz}-1
\end{array}\right|}=0
\end{displaymath}
Since we are employing the constrain that there can not be any doubly occupied 
sites so considering $U\rightarrow\infty$, our vanishing determinant be as 
follows:
\begin{equation}
{\left |\begin{array}{cccc}
I_{0} & -2JI_{x} & -2JI_{y} & -2JI_{z}\\
I_{x} & -2JI_{xx}-1 & -2JI_{xy} & -2JI_{zx}\\
I_{y} & -2JI_{xy} & -2JI_{yy}-1 & -2JI_{yz}\\
I_{z} & -2JI_{zx} & -2JI_{yz} & -2JI_{zz}-1
\end{array}\right|}=0
\end{equation}
Now because of lattice symmetry we can write 
\begin{center}
$I_{x}=I_{y}=I_{z}$\\$I_{xx}=I_{yy}=I_{zz}$\\$I_{xy}=I_{yz}=I_{zx}$
\end{center}
Also these lattice integrals can be written as:
\begin{equation}
I_{x}=\frac{1}{3}(\frac{1}{4t}-\frac{E}{4t}I_{0})
\end{equation}
\begin{equation}
I_{xx}=-\frac{E}{4t}I_{x}-2I_{xy}
\end{equation}
So we need to find out $I_{0}$ and $I_{xx}$ or $I_{xy}$.\\

{\bf Solution of Lattice Integrals}\\
Previously mentioned lattice integrals are commonly known as lattice Green's 
function. The most general representation of lattice green's function is,\cite{Hor}
\begin{displaymath}
G(s,l,m,n)=\frac{1}{\pi^{3}}\int_{0}^{\pi}\int_{0}^{\pi}\int_{0}^{\pi}\frac{\cos{lx}\cos{my}\cos{nz}dxdydz}{s-(\cos{x}+\cos{y}+\cos{z})}
\end{displaymath}
\begin{displaymath}
G(s)=G(s,0,0,0)=\frac{1}{\pi^{3}}\int_{0}^{\pi}\int_{0}^{\pi}\int_{0}^{\pi}\frac{dxdydz}{s-(\cos{x}+\cos{y}+\cos{z})}
\end{displaymath}
This integral defines a single valued analytic function G(s) in the complex 
s-plane cut along the real axis from -3 to +3. in most physical applications 
one usually requires the limiting behavior of the Green function G(s) as s 
approaches the real axis\cite{Joy}.\\
Another way we can write as\cite{Joyce} 
\begin{displaymath}
\frac{1}{3}G(s)=\frac{1}{3}\frac{1}{\pi^{3}}\int_{0}^{\pi}\int_{0}^{\pi}\int_{0}^{\pi}\frac{dxdydz}{s-\frac{1}{3}(\cos{x}+\cos{y}+\cos{z})}
\end{displaymath}
\begin{equation}
G_{1}(s)=\frac{1}{\pi^{3}}\int_{0}^{\pi}\int_{0}^{\pi}\int_{0}^{\pi}\frac{dxdydz}{s-\frac{1}{3}(\cos{x}+\cos{y}+\cos{z})}
\end{equation}
The series representation of the Green's function is\cite{Joyce}:
\begin{equation}
G_{j}(s)=\frac{1}{s}\sum_{n=0}^{\infty}\frac{p_{n}^{(j)}}{s^{n}}     
\end{equation}
where j=1, 2, 3, $1\leq|s|<\infty$ and
\begin{displaymath}
p_{n}^{(j)}=\frac{1}{\pi^{3}}\int_{0}^{\pi}\int_{0}^{\pi}\int_{0}^{\pi}[\lambda_{j}(\theta_{1},\theta_{2},\theta_{3})]^{n}d\theta_{1}d\theta_{2}d\theta_{3}
\end{displaymath}
Now for simple cubic lattice $\lambda_{1}(\theta_{1},\theta_{2},\theta_{3})=\frac{1}{3}(\cos\theta_{1}+\cos\theta_{2}+\cos\theta_{3})$
The recurrence relation for simple cubic lattice coefficient $p_{2n}^{1}$ is
\begin{equation}
36(n+1)^{3}p_{2n+2}^{(1)}-2(2n+1)(10n^{2}+10n+3)p_{2n}^{(1)}+n(4n^{2}-1)p_{2n-2}^{(1)}=0
\end{equation}
from eq. (52) and (53) $G_{1}(s)$ can be written as a solution of the linear third order differential equation
\begin{equation}
(s^{2}-1)(9s^{2}-1)\frac{d^{3}G_{1}}{ds^{3}}+6s(9s^{2}-5)\frac{d^{2}G_{1}}{ds^{2}}+3(21s^{2}-4)\frac{dG_{1}}{ds}+9sG_{1}=0
\end{equation}
So $G_{1}(s)$ can be written as,
\begin{equation}
G_{1}(s)=\frac{1-9\xi^{4}}{s(1-\xi)^{3}(1+3\xi)}[\frac{2}{\pi}\kappa(k_{1})]^{2}
\end{equation}
where,
\begin{displaymath}
k_{1}=\sqrt{\frac{16\xi^{3}}{(1-\xi)^{3}(1+3\xi)}}
\end{displaymath}
\begin{displaymath}
\xi=\xi(s)=\Bigg(1+\sqrt{1-\frac{1}{s^{2}}}\Bigg)^{-1/2}\Bigg(1-\sqrt{1-\frac{1}{9s^{2}}}\Bigg)^{1/2}
\end{displaymath}
and $\kappa(k_{1})$ is the complete elliptic integral of the first kind.\\
Also 
\begin{equation}
G_{1}(1)=3(18+12\sqrt{2}-10\sqrt{3}-7\sqrt{6})[\frac{2}{\pi}\kappa((2-\sqrt{3})(\sqrt{3}-\sqrt{2}))]^{2}
\end{equation}
So from this knowledge with proper adjustment of 
coefficients we can write:
\begin{equation}
I_{0}=\frac{1}{12t}G_{1}(s)
\end{equation}
Here we like to mention that elliptic integrals 
diverges for modulus of 1, the physical significance of the divergence in our 
work is the electrons are on the energy band therefore in our earlier analysis 
we had taken care of that divergence keeping our calculation just below the 
band, but here when we substitute our parameters as electrons are on the 
energy band we found the modulus is not 1 therefore there is no any sharp 
divergence of elliptic integral rather there is a local maxima, 
and we are returning with a numerical value of the integral on the band. Its 
a striking feature of the elliptic integral for simple cubic lattice. So from 
this knowledge we can evaluate $I_{0}$. But one of the integral among $I_{xx}$ 
or $I_{xy}$ are still to be evaluated. To evaluate our integrals we had taken 
help from the recurrence relation of the Green function. With the knowledge of 
the recurrence relation for fcc lattice\cite{Mic} and triangular lattice\cite{Hori}, we have constructed the recurrence relation for simple cubic lattice. For 
nearest neighbour (6 nn for S.C) interaction only the recurrence is as follows:
\begin{displaymath}
G(l+1,m,n)+G(l-1,m,n)+G(l,m+1,n)+G(l,m-1,n)+G(l,m,n+1)+G(l,m,n-1)
\end{displaymath}
\begin{equation}
=2\delta_{l0}\delta_{m0}\delta_{n0}-2sG(l,m,n)
\end{equation}   
where,
\begin{displaymath}
G(l+1,m,n)=\frac{1}{\pi^{3}}\int_{0}^{\pi}\int_{0}^{\pi}\int_{0}^{\pi}\frac{\cos{(l+1)x}\cos{my}\cos{nz}}{s+(\cos{x}+\cos{y}+\cos{z})}dxdydz
\end{displaymath}
and so on,
Let l=1, m=0, n=0 then
\begin{displaymath}
G(2,0,0)+G(0,0,0)+G(1,1,0)+G(1,-1,0)+G(1,0,1)+G(1,0,-1)=-2sG(1,0,0)
\end{displaymath}
Due to the symetric structure of the lattice
$G(1,1,0)=G(1,-1,0)=G(1,0,1)=G(1,0,-1)$
Hence, $G(2,0,0)+G(0,0,0)+4G(1,1,0)+2sG(1,0,0)=0$\\
Adjusting the coefficients for our calculation,
\begin{equation}
I_{xx}=-\frac{E}{4t}I_{x}-2I_{xy}
\end{equation}
where,
$I_{xx}=\frac{1}{8t}(G(2,0,0)+G(0,0,0))$, $I_{0}=\frac{1}{4t}G(0,0,0)$, $I_{x}=\frac{1}{4t}G(1,0,0)$
Solving eq. (50) and (59) $I_{xy}$ is determined.
\begin{equation}
I_{xy}=-\frac{E}{8t}I_{x}-\frac{1}{2}I_{0}
\end{equation}
Let us now summaries the lattice integral values:
\begin{displaymath}
I_{0}=\frac{1}{12t}G_{1}(s), s=\frac{E}{4t}
\end{displaymath}
\begin{displaymath}
I_{x}=\frac{1}{3}(\frac{1}{4t}-\frac{E}{4t}I_{0})
\end{displaymath}
\begin{equation}
I_{xy}=-\frac{E}{8t}I_{x}-\frac{1}{2}I_{0}
\end{equation}
\begin{displaymath}
I_{xx}=-\frac{E}{4t}I_{x}-2I_{xy}
\end{displaymath}
Substituting eq. (60) in eq. (48) and solving for J, the values are obtained as
{\bf 2}, {\bf 2}, {\bf 7.88}.
\section{RESULTS AND DISCUSSION:}
In our entire work we were involved in finding out the critical values of J 
($J_{c}$) for which bounstate formation of two electron system is possible, and
our results obtained are as follows:
\begin{center}
\begin{tabular}{|l|l|}
\hline
Lattice Type & $J_{c}$ Values \\
\hline
One Dimensional Chains & 2t\\
\hline
Two Dimensional Square Lattice &2t, 7.32t\\
\hline
Two Leg Ladder & 2t\\
\hline
Three Dimensional Cubic Lattice & 2t, 2t, 7.88t\\
\hline
\end{tabular}
\end{center}
Also we have plotted $J_{c}/t$ Vs. $|E_{bs}|/t$ for one dimensional chain, two 
dimensional square lattice and two leg ladder.\\
For one dimensional chain like system has been already solved for 
J=2t\cite{P,Ex}.\\
For square lattice the critical value of J=2t can seen in papers of Lin \cite{Em,Lin}.
In our work we are proposing that for two leg ladder lattice the critical value
of J is 2t, and it is quite desirable because two leg ladder is a simplified 
form of n leg ladder system which is nothing but a 2 dimensional system.\\
We are also proposing for the first time that for three dimensional lattice the
 critical value of J is 4. \\
Let us now try to understand the significance of the multiple value of two 
dimensional and three dimensional lattice for the basis of group theory. \\
{\bf Square Lattice}\\
We know that the symmetries of a square lattice is represented by $C_{4V}$ 
group. The character table for $C_{4V}$ group is as follows:\\
\begin{center}
\begin{tabular}{l|l|l|l|l|l|l|l|l}
\hline
Characters &E&$C_{4}$&$C_{4}^{2}$&$C_{4}^{3}$&$m_{x}$&$m_{y}$&$\sigma_{u}$&$\sigma_{v}$ \\
\hline
$\chi^{(1)}$ &1&1&1&1&1&1&1&1\\
\hline
$\chi^{(2)}$ &1&-1&1&-1&-1&-1&1&1\\
\hline
$\chi^{(3)}$ &1&-1&1&-1&1&1&-1&-1\\
\hline
$\chi^{(4)}$ &1&1&1&1&-1&-1&-1&-1\\
\hline
$\chi^{(5)}$ &2&0&-2&0&0&0&0&0\\
\hline
\end{tabular}
\end{center}
From the character table we can find out the irreducible representation of the 
group and the corresponding basis functions\cite{Jo}.
\begin{center}
\begin{tabular}{l|l}
\hline
Irreducible Representation & Basis Functions \\
\hline
$\Gamma^{(1)}$&1\\
\hline
$\Gamma^{(2)}$&$xy$\\
\hline
$\Gamma^{(3)}$&${x}^{2}-{y}^{2}$\\
\hline
$\Gamma^{(4)}$&$xy({x}^{2}-{y}^{2})$\\
\hline
$\Gamma^{(5)}$&$(x,y)$\\
\hline
\end{tabular}
\end{center}
The possible pairing symmetries correspond to these irreducible representation 
and to the basis functions. In case of square crystals the singlet orders are 
called $s$, $d_{x^{2}-y^{2}}$, $d_{xy}$ and $g$. The corresponding basis 
functions are $1$, ${x}^{2}-{y}^{2}$, $xy$, $xy({x}^{2}{y}^{2})$. It is 
reasonable to classify order parameters as "s-wave like" and "d-wave-like". So 
in our analysis J=2t corresponds to s-wave and J=7.32t corresponds to d-wave\cite{Sup}.\\
{\bf Simple Cubic Lattice}\\
It is known that the simple cubic lattice corresponds to $O_{h}$ group.
The character table for $O_{h}$ group is as follows\cite{Co}:
\begin{center}
\begin{tabular}{l|l|l|l|l|l|l|l|l|l|l}
\hline
Characters &E&$8C_{3}$&$6C_{2}$&$6C_{4}$&$3C_{2}(=C_{4}^{2})$&$i$&$6S_{4}$&$8S_{6}$&$3\sigma_{h}$&$6\sigma_{d}$ \\
\hline
$A_{1g}$ &1&1&1&1&1&1&1&1&1&1\\
\hline
$A_{2g}$ &1&1&-1&-1&1&1&-1&1&1&-1\\
\hline
$E_{g}$  &2&-1&0&0&2&2&0&-1&2&0\\
\hline
$T_{1g}$ &3&0&-1&1&-1&3&1&0&-1&-1\\
\hline
$T_{2g}$ &3&0&1&-1&-1&3&-1&0&-1&1\\
\hline
$A_{1u}$ &1&1&1&1&1&-1&-1&-1&-1&-1\\
\hline
$A_{2u}$ &1&1&-1&-1&1&-1&1&-1&-1&1\\
\hline
$E_{u}$  &2&-1&0&0&2&-2&0&1&-2&0\\
\hline
$T_{1u}$ &3&0&-1&1&-1&-3&-1&0&1&1\\
\hline
$T_{2g}$ &3&0&1&-1&-1&-3&1&0&1&-1\\
\end{tabular}
\end{center}
So the irreducible representation and basis functions for singlet pairing are\cite{Sup}\\
\begin{center}
\begin{tabular}{l|l}
\hline
Irreducible Representation & Basis Functions \\
\hline
$\Gamma_{1}{(A_{1g})}$&1\\
\hline
$\Gamma_{2}{(A_{2g})}$&$(k_{x}^{2}-k_{y}^{2})(k_{y}^{2}-k_{z}^{2})(k_{z}^{2}-k_{x}^{2})$\\
\hline
$\Gamma_{3}{(E_{g})}$&$2k_{z}^{n}-k_{x}^{n}-k_{y}^{n}, k_{x}^{n}-k_{y}^{n}$\\
$$&$n=2; 4$\\
\hline
$\Gamma_{4}{(T_{1g})}$&$k_{x}k_{y}(k_{x}^{n}-k_{y}^{n}),k_{y}k_{z}(k_{y}^{n}-k_{z}^{n}),k_{z}k_{x}(k_{z}^{n}-k_{x}^{n})$\\
$$&$n=2; 4; 6$\\
\hline
$\Gamma_{5}{(T_{2g})}$&$k_{x}k_{y}k_{z}^{n},k_{y}k_{z}k_{x}^{n},k_{z}k_{x}k_{y}^{n}$\\
$ $&$n=0;2;4$\\
\hline
\end{tabular}
\end{center}
So in a cubic lattice, $\Gamma_{1}(A_{1g})$ represents s-wave pairing, 
$\Gamma_{3}(E_{g})$ (for n=2) and $\Gamma_{5}(T_{2g})$ (for n=0) represents 
d-wave pairing\cite{Sup}. Also $\Gamma_{3}(E_{g})$ is doubly degenerate. 
So from the evaluated value of of J, we can conclude that J=2t value corresponds to $\Gamma_{3}(E_{g})$ representation and J=7.88t corresponds to 
$\Gamma_{1}(A_{2g})$ representation. 
\section{CONCLUSION AND FUTURE PROSPECTS:}
The analytical result of J for simple cubic lattice is a new result. 
The result can be cross examined by bilayer consideration of the three 
dimensional cubic lattice. The bilayer consideration is analogous with the two leg ladder consideration. Physically it can be vilualised that, n number of layers togather is forming a three dimensional structure provided the separation between the layers is sufficiently small. So starting the calculation for bilayer one can extend the calculation to n number of layer which can give the similar result as of the three dimension. A natural extrapolation can be investigating superconductivity in these lattices and studying the phase diagram.   

\end{document}